# Building a GPU-Accelerated Multivariate Statistics Platform

Mike Crowhurst
Graduate Student, Dr. Bing Zhang Department of Statistics
University of Kentucky

### *Abstract*

Classical multivariate statistical methods, such as covariance estimation and principal component analysis, are mathematically well established, but applying them at extremely large scale remains challenging. When datasets grow to billions of observations, the limiting factors are no longer arithmetic operations but practical issues such as data movement, input–output overhead, and numerical stability. In particular, repeated passes over text-based data and the presence of non-random identifier variables can dominate runtime and introduce subtle numerical problems.

This paper presents an end-to-end case study of scaling these familiar methods on a single multi-GPU system. Using C++ and CUDA, I developed and validated a GPU-based workflow that emphasizes careful data representation, validation against known invariants, efficient computation of sufficient statistics, and attention to numerical correctness. By computing column sums and cross-product matrices in a single pass over a 10-billion-row dataset, the workflow supports downstream estimation of means, covariance, correlation, and principal components without revisiting the raw data.

The contribution of this work is not a new statistical method, but a set of practical design choices and lessons learned from pushing standard techniques into a regime where they are not typically applied. The results underscore the dominant role of data movement, the value of validation strategies grounded in known identities, and the importance of numerically stable computation at scale.

### *Introduction*

Working with datasets on the order of billions of records introduces challenges that go well beyond raw computational throughput. At this scale, the dominant costs are often associated with data movement: transferring data between storage and memory, moving data from host memory to accelerator devices, and repeatedly parsing text-based formats (Reed, 2015). These overheads can easily overwhelm available compute resources, leaving even powerful hardware underutilized. At the same time, achieving reliable numerical results becomes more difficult as the number of observations grows, particularly when large accumulations are performed in finite-precision arithmetic.

Classical multivariate methods—such as covariance estimation and principal component analysis—remain mathematically sound and conceptually appropriate for large datasets. However, in practice, they can become difficult to apply efficiently at extreme scale (Fan, 2018); (Demmel, 2012). Much of this difficulty arises not from the statistical methods themselves, but from how the data are handled. Repeated passes over raw data, reliance on text-based formats such as comma-separated values (CSV), and the inclusion of identifier variables (for example, line numbers) can introduce substantial performance penalties and, in some cases, subtle numerical pathologies.

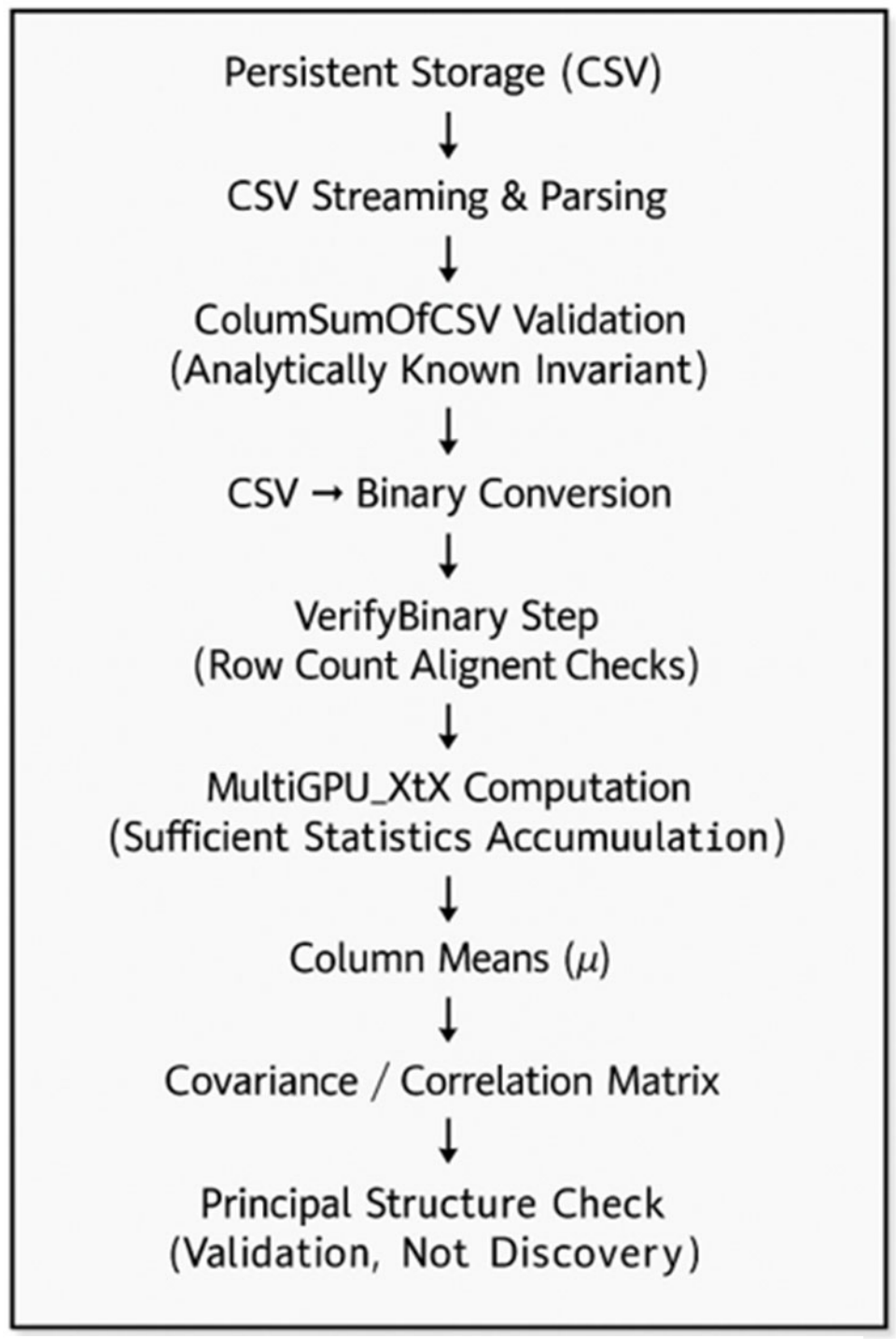

**Figure 1: Pipeline Process**

This paper presents an end-to-end case study of scaling classical multivariate computations to a dataset (see Table 1) containing 10 billion observations. Using a single accelerator-rich node equipped with five GPUs, I developed and validated a GPU-based workflow implemented in C++ and CUDA. The approach emphasizes careful data representation, validation against known invariants, single-pass computation of sufficient statistics, and attention to numerical correctness (see Figure 1).

The intent of this study is not to introduce new statistical methods, but to document practical design decisions and lessons learned when applying standard techniques in a regime where they are not typically used. In this setting, issues such as data movement, validation, and numerical stability become central, and addressing them effectively is essential for obtaining both efficient and trustworthy results.

### *Background and Motivation*

Classical multivariate techniques such as covariance estimation and principal component analysis (PCA) are foundational tools in data analysis (Anderson, 2003); (Mardia, 1979); (Pearson, 1901). In many applications, the number of variables is modest, and these methods are straightforward to apply. However, when the number of observations becomes extremely large, the primary challenges shift. At that point, the difficulty is no longer in the mathematics itself, but in the cost of accessing, moving, and repeatedly processing the data (Fan, 2018).

Traditional workflows often rely on multiple passes over text-based input formats and CPU-centric processing. While this is adequate at smaller scales, it becomes increasingly inefficient as datasets grow into the billions of records—for example, in web crawl or large corporate datasets. Input–output overhead, data parsing, and memory transfers can dominate execution time, leaving accelerator hardware underutilized. At the same time, numerical issues that are negligible in smaller settings—such as loss of precision in large accumulations or the unintended influence of deterministic identifier variables—can begin to affect results in meaningful ways if they are not handled carefully.

***Table 1: Dataset used in this study.***

| Column | Description | Formula |
|---|---|---|
| 1 | Identifier | A++ |
| 2 | Random Number Between 3 and 8 | int B = randBetween(3, 8); |
| 3 | Random Number Between 1 and 10 | int C = randBetween(1, 10); |
| 4 | Random Number Between 1 and 100 | int D = randBetween(1, 100); |
| 5 | Log of C / Log of B * 100 | double E = int(log(C) / log(B) * 100); |
| 6 | Log of D / Log of B * 10000 | double F = round((log(D) / log(B)) * 10000); |
| 7 | Integer of the Absolute Value of the Cosine of C * 100 | double G = int(abs(cos(C)) * 100); |
| 8 | Integer of the Absolute Value of the Sine of D * 100 | double H = int(abs(sin(D)) * 100); |
| 9 | One divided by the tangent of C | double cotC = 1.0 / tan(C); |
| 10 | Absolute value of the Cotangent of C * 1000 Rounded | double I = round(abs(cotC) * 1000); |
| 11 | Absolute value of the Tangent of D | double J = abs(tan(D)); |
| 12 | D / C (Integer Division) | int K = D / C; |

These limitations motivate a different approach, one that focuses on minimizing data movement and organizing computation around sufficient statistics. Sufficient statistics (Fisher, 1922); (Casella, 2002) summarize the information in the data relevant to a parameter of interest, so that once they are known, the full dataset no longer needs to be revisited. In this setting, reducing a large dataset to compact summaries in a single pass effectively decouples the cost of downstream analysis from the size of the raw data.

When combined with accelerator-rich systems, this perspective provides a practical path for scaling classical statistical methods while maintaining numerical correctness. The work presented here follows this approach, emphasizing an end-to-end workflow that reflects these considerations in a realistic computing environment.

### ***System Overview***

The system used in this study was a single-node, accelerator-rich configuration consisting of five NVIDIA GeForce RTX™ 3070 GPUs (NVIDIA, Corporation, 2026) connected through a standard PCIe-based motherboard (Micro-Star International (MSI), 2026), with 32 gigabytes of system memory. A 4-terabyte solid-state drive (SSD) was used to store the full dataset and intermediate results locally. This setup reflects a class of readily available workstation or small-server systems that are commonly used for data-parallel workloads in contemporary high-performance computing environments.

The system ran Microsoft Windows 11, with development carried out in Visual Studio 2022. CUDA version 12.8 was used to enable the use of multiple GPUs from C++ programs. All computations were performed within a single node, avoiding distributed-memory communication. This design choice was intentional, allowing the study to focus specifically on data movement, validation, and numerical behavior within an accelerator-rich environment, without introducing the additional complexity of a distributed system.

More broadly, the system was configured as a practical platform for end-to-end evaluation of data ingestion, validation, and the computation of sufficient statistics. By keeping the architecture relatively simple, it becomes easier to isolate workflow and numerical issues that would otherwise be obscured in a more complex distributed setting.

## **Validation and Infrastructure Testing**

At extreme data scales, errors in data ingestion, device coordination, or numerical accumulation can be difficult to detect—especially once performance optimizations are introduced (Reed, 2015); (Cebrian, 2020). In multi-GPU environments, incorrect results may arise from relatively subtle issues, such as skipped data, duplicated processing, race conditions in reductions, or incorrect partitioning of work across devices. With datasets containing billions of records, these problems are unlikely to be caught through ad hoc inspection or spot checks.

For this reason, the workflow in this study follows a validation-first approach, prioritizing correctness and infrastructure verification before any optimization for performance. The goal of this phase was to establish that the data pipeline, GPU orchestration, and reduction logic behaved as expected at scale before extending the workflow to full multivariate computation.

As an initial validation step, a GPU-accelerated program was developed to compute the sum of the first column of the dataset. This column contained a simple index variable (ranging from 1 to 10 billion) with a known analytical sum, providing an unambiguous check of end-to-end correctness.

The program streamed CSV data in fixed-size chunks, parsed only the first column, and distributed values across multiple GPUs for parallel reduction. Because the expected result was known in advance, any discrepancy immediately pointed to an error in parsing, partitioning, or reduction logic. This step made it possible to verify that every record was processed exactly once and that no data were skipped or duplicated during streaming.

Once the ingestion and reduction pipeline had been validated, the dataset was converted from a text-based CSV representation to a binary format. This conversion eliminated repeated parsing overhead and enabled more efficient streaming into accelerator memory. However, while binary formats improve performance, they also introduce the risk that errors during conversion may go unnoticed in later stages of analysis.

To address this, an additional validation step was performed after conversion. The binary file was checked against expected properties, including file size, row count, and column layout based on the known dataset dimensions. Records from the beginning, middle, and end of the file were also sampled and compared to the original CSV source to confirm numerical consistency and correct alignment of columns. These checks were intended to ensure that all observations were converted exactly once and that no corruption occurred during the transformation.

This validation confirmed that the binary representation faithfully preserved the original dataset and could be used reliably for subsequent computation of sufficient statistics. By verifying the conversion step explicitly, the workflow avoids attributing downstream numerical or performance issues to what might otherwise be undetected data integrity problems.

The column-sum program itself used a straightforward data-parallel reduction across the available GPUs. Each device was assigned a disjoint subset of the input data, performed reductions independently, and returned partial sums to the host, where they were accumulated into a final result.

Keeping this computation deliberately simple allowed the validation to focus on infrastructure behavior rather than algorithmic complexity. By reducing the task to a single scalar quantity with a known value, it became possible to isolate and verify multi-GPU coordination without introducing additional sources of error. Successful completion of this step demonstrated that multiple GPUs could be driven concurrently from a single host process and that device-level reductions produced consistent and correct results.

This validation phase led to several observations that informed the remainder of the study. First, it confirmed that data movement and parsing dominated execution time, while GPU computation itself was comparatively fast—reinforcing the need to minimize repeated passes over raw data. Second, it highlighted the value of using analytically verifiable results when validating large-scale workflows, particularly in settings where traditional debugging is impractical. Finally, it provided confidence that the underlying infrastructure was functioning correctly, allowing subsequent stages to focus on computing sufficient statistics and higher-order analyses. By isolating correctness early, the workflow avoids conflating performance behavior with potential data integrity or numerical errors.

**Data Representation and Sufficient Statistics**

Text-based formats such as comma-separated values (CSV) are widely used because they are simple and portable. However, at very large scales, they introduce substantial overhead. When datasets reach billions of records, the repeated parsing of text, string conversion, and delimiter handling can dominate execution time and place sustained pressure on the memory subsystem (Reed, 2015). These costs are incurred regardless of how simple the downstream numerical computations may be, and they can lead to significant underutilization of GPU hardware.

In preliminary experiments, CSV ingestion accounted for a disproportionate share of total runtime, even when only a subset of columns was used. This behavior motivated a shift away from text-based representations toward a format better suited for large-scale numerical computation. To address the inefficiency of repeated parsing, the dataset was converted to a fixed-width binary representation. Each record was stored in row-major order using double-precision values, allowing direct and contiguous access from both host and device memory. This eliminates parsing overhead and enables efficient streaming into GPU memory without intermediate string processing.

The binary format also provides a more stable foundation for computation, since the cost of accessing individual values becomes predictable and independent of any textual encoding. Once conversion and validation were complete, the binary dataset served as the canonical representation for all subsequent stages of the workflow.

Rather than making multiple passes over the full dataset for each statistical quantity, the workflow was organized around computing sufficient statistics in a single streaming pass. In particular, column sums and the cross-product matrix $X^\top X$were accumulated as the data were streamed through the system (see Table 2).

These quantities are sufficient to compute column means, covariance matrices, and correlation matrices. Once they are available, the remaining analysis no longer requires access to the full dataset. The computations were carried out in a data-parallel manner across multiple

*Table 2: Covariance Matrix*

| 1.00000000000000 | 0.00000000167067 | 0.00000002384430 | 0.00000347052000 | 0.00000027395700 | 0.00000000299095 | 0.00000000299085 | 0.00000027395000 | 0.00000347057000 | 0.00000002384450 |
|---|---|---|---|---|---|---|---|---|---|
| 0.00000000167067 | 1.00000000000000 | 0.00000000167087 | 0.00000002384450 | 0.00000347057000 | 0.00000027396100 | 0.00000000299076 | 0.00000000299083 | 0.00000027397100 | 0.00000347042000 |
| 0.00000002384430 | 0.00000000167087 | 1.00000000000000 | 0.00000000167080 | 0.00000002384340 | 0.00000347049000 | 0.00000027396200 | 0.00000000299098 | 0.00000000299098 | 0.00000027398300 |
| 0.00000347052000 | 0.00000002384450 | 0.00000000167080 | 1.00000000000000 | 0.00000000167003 | 0.00000002384400 | 0.00000347043000 | 0.00000027395100 | 0.00000000299108 | 0.00000000299096 |
| 0.00000027395700 | 0.00000347057000 | 0.00000002384340 | 0.00000000167003 | 1.00000000000000 | 0.00000000167032 | 0.00000002384370 | 0.00000347033000 | 0.00000027397600 | 0.00000000299082 |
| 0.00000000299095 | 0.00000027396100 | 0.00000347049000 | 0.00000002384400 | 0.00000000167032 | 1.00000000000000 | 0.00000000167069 | 0.00000002384390 | 0.00000347049000 | 0.00000027395800 |
| 0.00000000299085 | 0.00000000299076 | 0.00000027396200 | 0.00000347043000 | 0.00000002384370 | 0.00000000167069 | 1.00000000000000 | 0.00000000167084 | 0.00000002384420 | 0.00000347060000 |
| 0.00000027395000 | 0.00000000299083 | 0.00000000299098 | 0.00000027395100 | 0.00000347033000 | 0.00000002384390 | 0.00000000167084 | 1.00000000000000 | 0.00000000167096 | 0.00000002384390 |
| 0.00000347057000 | 0.00000027397100 | 0.00000000299098 | 0.00000000299108 | 0.00000027397600 | 0.00000347049000 | 0.00000002384420 | 0.00000000167096 | 1.00000000000000 | 0.00000000167104 |
| 0.00000002384450 | 0.00000347042000 | 0.00000027398300 | 0.00000000299096 | 0.00000000299082 | 0.00000027395800 | 0.00000347060000 | 0.00000002384390 | 0.00000000167104 | 1.00000000000000 |

GPUs, with each device processing a disjoint subset of the data and contributing partial results that were subsequently combined. By limiting access to the full dataset to a single pass, the workflow minimizes data movement and avoids repeated traversal of the binary file, which would otherwise reintroduce I/O and memory bottlenecks.

After the sufficient statistics were computed, all subsequent analyses—including mean estimation, covariance and correlation computation, and principal component analysis—were performed using only the aggregated matrices.

This separation between data ingestion and statistical analysis provides two key benefits. First, it allows the sufficient statistics to be reused across multiple analyses without additional data movement. Second, it simplifies numerical reasoning by confining large-scale accumulation to a single, well-defined step, reducing the risk of cumulative numerical error from repeated processing. In this way, sufficient statistics serve as the boundary between large-scale data handling and smaller-scale numerical analysis, making it feasible to apply classical statistical methods efficiently and reliably at data sizes that would otherwise be impractical.

## Numerical Considerations

When computations involve extremely large numbers of observations, numerical effects that are negligible at smaller scales can become significant. In particular, the accumulation of large sums in finite-precision arithmetic can lead to catastrophic cancellation, where subtracting nearly equal large quantities results in a substantial loss of precision. In the context of covariance computation, this issue arises when quantities such as $X^\top X$and $N\,\mu\mu^\top$are formed independently and then combined.

At large $N$, both terms can be several orders of magnitude larger than their difference, which amplifies floating-point round-off error. These effects do not indicate a failure of the underlying algorithms; rather, they reflect the interaction between scale and numerical representation. As a result, workflows that are mathematically correct can still produce unstable or misleading results when applied at extreme scale.

A related issue encountered in this study arose from the inclusion of a deterministic identifier variable. The first column of the dataset consisted of a sequential index used for identification and validation purposes. When this column was included in covariance and correlation calculations, it introduced a large deterministic trend that dominated the numerical scale of the computation.

Because this identifier is not random, its variance grows quadratically with the number of observations, overwhelming the contributions of the genuinely random variables. In practice, this led to negative variance estimates and undefined correlation values due to catastrophic cancellation during matrix operations. Importantly, these failures were not properties of the simulated data themselves, but artifacts of including a non-random identifier in the analysis.

To address this issue, the identifier column was excluded from all covariance, correlation, and principal component computations. Once removed, the remaining variables produced stable variance estimates, well-defined correlation matrices, and consistent results across repeated runs. This correction restored numerical stability without changing the underlying statistical methodology, and it highlights the importance of explicitly distinguishing between identification fields and random variables in large-scale workflows.

These observations reinforce an important distinction between computational speed and computational correctness in high-performance environments. At extreme scale, correctness cannot be assumed to follow automatically from mathematical formulation or algorithmic efficiency. Instead, careful attention must be paid to numerical conditioning, data semantics, and the order of operations.

By explicitly accounting for catastrophic cancellation and the treatment of deterministic variables, this workflow shows how classical statistical methods can be applied reliably at large scale. These considerations are essential to ensuring that performance gains from accelerator-rich systems do not come at the expense of numerical validity.

**Principal Component Analysis Results**

Principal component analysis (PCA) was performed using the covariance matrix derived from the sufficient statistics computed in the previous stage. Because the covariance matrix was constructed from column sums and the cross-product matrix X⊤XX^\top XX⊤X, no additional access to the original dataset was required. In this workflow, all large-scale data movement is confined to the sufficient-statistics phase, allowing PCA to be carried out entirely on relatively small, dense matrices.

The eigenvalues and eigenvectors of the covariance matrix were computed using standard numerical linear algebra routines. Given that the number of variables was small relative to the number of observations, the computational cost of PCA itself was negligible compared to data ingestion and the accumulation of sufficient statistics (see Table 3).

The resulting eigenvalue spectrum was nearly uniform, with each principal component accounting for approximately the same proportion of total variance. No individual component, or small subset of components, captured a dominant share of the variance.

This behavior is consistent with how the dataset was constructed. The variables were generated independently and without any imposed correlation structure, so the covariance matrix is expected to be approximately diagonal (after appropriate normalization), with eigenvalues of similar magnitude.

***Table 3 – principal components results***

|  | |  | |  | |  |
|---|---|---|---|---|---|---|
| PC1: | Eigenvalue = | 1.0000100 | Variance % = | 10.00010 | Cumulative % = | 10.0001 |
| PC2: | Eigenvalue = | 1.0000100 | Variance % = | 10.00010 | Cumulative % = | 20.0001 |
| PC3: | Eigenvalue = | 1.0000000 | Variance % = | 10.00000 | Cumulative % = | 30.0002 |
| PC4: | Eigenvalue = | 1.0000000 | Variance % = | 10.00000 | Cumulative % = | 40.0002 |
| PC5: | Eigenvalue = | 1.0000000 | Variance % = | 10.00000 | Cumulative % = | 50.0002 |
| PC6: | Eigenvalue = | 0.9999990 | Variance % = | 9.99999 | Cumulative % = | 60.0002 |
| PC7: | Eigenvalue = | 0.9999970 | Variance % = | 9.99997 | Cumulative % = | 70.0002 |
| PC8: | Eigenvalue = | 0.9999960 | Variance % = | 9.99996 | Cumulative % = | 80.0001 |
| PC9: | Eigenvalue = | 0.9999940 | Variance % = | 9.99994 | Cumulative % = | 90.0001 |
| PC10: | Eigenvalue = | 0.9999930 | Variance % = | 9.99993 | Cumulative % = | 100.0000 |

**Loadings (columns = PCs):**

| | | | | | | | | | | |
|---|---|---|---|---|---|---|---|---|---|---|
| PC1: | -0.369674 | -0.343743 | -0.168364 | -0.371863 | -0.280267 | -0.280280 | -0.371864 | -0.168350 | -0.343757 | -0.369666 |
| PC2: | -0.322256 | 0.422066 | -0.230524 | -0.120121 | 0.387859 | -0.387864 | 0.120157 | 0.230513 | -0.422063 | 0.322269 |
| PC3: | -0.150309 | 0.218133 | 0.265120 | -0.461127 | 0.383294 | 0.383278 | -0.461123 | 0.265121 | 0.218104 | -0.150297 |
| PC4: | 0.422062 | -0.120095 | -0.387886 | 0.230524 | 0.322258 | -0.322265 | -0.230542 | 0.387857 | 0.120133 | -0.422048 |
| PC5: | 0.224933 | 0.381990 | -0.462966 | -0.268076 | -0.131428 | -0.131416 | -0.268051 | -0.463005 | 0.381984 | 0.224959 |
| PC6: | 0.230503 | 0.387867 | 0.422051 | 0.322268 | 0.120143 | -0.120120 | -0.322234 | -0.422087 | -0.387864 | -0.230543 |
| PC7: | 0.445985 | -0.121954 | 0.352778 | -0.257237 | -0.309198 | -0.309210 | -0.257269 | 0.352746 | -0.121998 | 0.445966 |
| PC8: | 0.120134 | -0.230552 | 0.322251 | -0.387867 | 0.422070 | -0.422056 | 0.387864 | -0.322247 | 0.230522 | -0.120127 |
| PC9: | 0.302078 | -0.416500 | -0.250147 | -0.105071 | 0.402069 | 0.402070 | -0.105110 | -0.250135 | -0.416490 | 0.302099 |
| PC10: | 0.387879 | 0.322242 | -0.120133 | -0.422062 | -0.230517 | 0.230536 | 0.422060 | 0.120118 | -0.322265 | -0.387864 |

In many applied settings, PCA is used to identify latent structure or to reduce dimensionality. In this case, however, the absence of dominant components should not be viewed as a limitation of the method. Rather, it is the expected result for a dataset composed of independent variables with comparable variance.

The uniform eigenvalue spectrum also serves as an important consistency check for the workflow. It suggests that the statistical properties of the data were preserved throughout data ingestion, format conversion, multi-GPU reduction, and covariance computation. In particular, it indicates that no artificial correlations were introduced during processing.

For this reason, PCA was used here primarily as a validation tool rather than as a means of discovering structure. Because the statistical properties of the data were known in advance, the PCA results provide a high-level diagnostic of numerical correctness.

The absence of dominant components, together with stable eigenvalues across repeated runs, reinforces confidence in the correctness of the sufficient-statistics computation and subsequent numerical operations. In this sense, PCA complements earlier validation steps by providing a global summary of variance structure that would be sensitive to errors in covariance estimation or numerical instability.

More broadly, this use of PCA highlights an important distinction in large-scale workflows: statistical tools can serve not only analytical purposes, but also as effective validation mechanisms when used with an understanding of the underlying data.

**Discussion and Lessons Learned**

One of the most important lessons from this study was the dominant role of data representation in overall system behavior. Although GPU acceleration significantly reduced the cost of numerical computation, these gains were initially masked by the overhead of ingesting and parsing text-based CSV data. Repeated passes over CSV files introduced substantial latency and limited effective utilization of the available hardware.

The transition to a fixed-width binary representation had a far greater impact than any low-level optimization of computational kernels. Once parsing and representation overhead were removed, the system was able to stream data efficiently into GPU memory, allowing the numerical computations to proceed as intended. This reinforces a key point for large-scale workflows: data movement and representation often matter more than arithmetic optimization, and they need to be addressed early in the design process.

In contrast, fine-grained tuning of GPU kernels played a relatively minor role in overall performance. The reduction kernels used for column sums and cross-product accumulation were intentionally simple and did not rely on advanced optimization techniques. Even so, once data movement costs were controlled, kernel execution accounted for only a small fraction of total runtime.

This outcome highlights a common pitfall in high-performance computing: focusing too early on kernel-level optimization while overlooking system-level bottlenecks. For data-intensive workloads at extreme scale, improvements in arithmetic throughput are unlikely to have a meaningful impact unless the costs of data access and movement are addressed first.

The decision to implement the workflow on a single node with multiple GPUs involved deliberate trade-offs. On the one hand, the single-node design simplified development and debugging, eliminated distributed communication overhead, and made it possible to focus on data movement, validation, and numerical behavior within a controlled environment. For data-parallel tasks such as sufficient-statistics computation, this configuration proved both effective and scalable within the limits of the problem.

At the same time, the approach has clear limitations. It is constrained by the memory capacity and I/O bandwidth of a single system and does not address issues related to distributed data placement or inter-node communication. While the single-node configuration was sufficient for the scale considered here, extending the approach further would likely require a distributed implementation.

The workflow developed in this study is best suited to problems characterized by very large numbers of observations and relatively modest dimensionality, where sufficient statistics can be computed in a single pass and reused across analyses. Examples include large observational datasets, financial transaction records, healthcare data, educational assessments, simulation output, and other log-based data where repeated access to raw data is prohibitively expensive.

In contrast, the approach is less well suited to problems that require frequent global synchronization, iterative passes over the full dataset, or complex dependency structures that cannot be reduced to a single streaming pass. In such cases, distributed architectures or alternative computational strategies may be more appropriate.

More broadly, these results highlight the importance of matching computational strategy to problem structure. Recognizing when a single-pass, sufficient-statistics approach is appropriate—and when it is not—is essential for using high-performance computing resources effectively and avoiding unnecessary system complexity.

**Limitations and Future Work**

A primary limitation of this study is its focus on a single-node, multi-GPU system. This design choice was intentional, allowing the workflow to isolate and examine data movement, validation, and numerical behavior within an accelerator-rich environment. However, it also constrains the scope of the findings. While the results demonstrate that classical multivariate methods can be scaled effectively within a single node for large-$N$, moderate-$p$ problems, they do not address challenges related to distributed data placement, inter-node communication, or fault tolerance. Extending the workflow to multi-node systems would introduce additional complexities that were beyond the scope of the present study.

The dataset used in this work was synthetically generated to provide controlled statistical properties and analytically verifiable behavior. This was advantageous for validation and numerical analysis, but it limits direct generalization to real-world data. In practice, observational datasets often contain missing values, heterogeneous distributions, and more complex dependency structures. While the synthetic data allowed the workflow to be evaluated under well-understood conditions, an important next step is to examine how the system performs when applied to less idealized data.

The current implementation also does not include distributed execution, and therefore does not address scaling limits imposed by single-node memory capacity or I/O bandwidth. For datasets that exceed the capabilities of a single system, distributed approaches would be necessary. A natural extension of this work is to investigate how sufficient-statistics computation can be partitioned across nodes while maintaining numerical correctness.

Several additional directions for future work follow from these results. Incorporating data generators with controlled correlation structures would provide a more complete validation of covariance and PCA behavior under known dependencies. Applying the workflow to real-world datasets would offer practical insight into the effects of data quality, sparsity, and irregular distributions. Finally, extending the sufficient-statistics framework to multi-node environments would allow a direct evaluation of the trade-offs between intra-node acceleration and inter-node communication, and would provide a more complete understanding of scalability for large-$N$ statistical workloads.

***Conclusion***

This work presented an end-to-end case study of scaling classical multivariate statistical computation to extremely large datasets using a single accelerator-rich node. Rather than introducing new statistical methods or specialized hardware, the focus was on workflow design, validation, and numerical correctness when applying well-established techniques at a scale where they are not typically used. By organizing the computation around sufficient statistics and validating each stage of the data pipeline, the study shows that correctness can be maintained even when the number of observations grows into the billions.

One of the central findings is that data representation, validation, and numerical considerations play a far greater role in overall system behavior than low-level kernel optimization. While GPU acceleration reduces the cost of arithmetic, those benefits can easily be overshadowed by the overhead of data ingestion and parsing. The shift from text-based input to a validated binary representation, combined with a single-pass computation of sufficient statistics, proved to be essential. These choices made it possible to reuse results efficiently, minimize data movement, and maintain numerical stability throughout the workflow.

More broadly, the results demonstrate that classical statistical methods remain viable at extreme scale when supported by appropriate system design and careful computational practice. For data-intensive workloads characterized by large numbers of observations and moderate dimensionality, approaches that prioritize validation, data handling, and numerical robustness provide a practical path forward. As data volumes continue to grow across scientific and applied domains, the ability to align statistical methods with system-level constraints will become increasingly important.

***The author used an AI-based assistant to support programming tasks (e.g., explaining C++/CUDA constructs, suggesting code organization, and assisting with documentation). The AI system did not generate data, perform experiments, or determine results or conclusions. All computational work, statistical analysis, and interpretation were conducted and verified by the author.***